\documentclass{article}
\usepackage{arxiv}

\usepackage[utf8]{inputenc}
\usepackage[T1]{fontenc}
\usepackage[hidelinks]{hyperref}
\usepackage{url}
\usepackage{booktabs}
\usepackage{amsfonts}
\usepackage{amsmath}
\usepackage{amssymb}
\usepackage{nicefrac}
\usepackage{microtype}
\usepackage{graphicx}
\usepackage{natbib}
\usepackage{caption}
\usepackage{subcaption}
\usepackage{float}
\usepackage{orcidlink}

\setkeys{Gin}{keepaspectratio}

\newcommand{\runninghead}{Joint Velocity and Impedance Inversion via Diffusion Models}

\providecommand{\keywords}[1]{%
  \vspace{0.6em}\noindent\textbf{Keywords:}~#1\vspace{0.4em}}

\title{Bayesian Joint Velocity and Impedance Inversion via Diffusion Models
Conditioned on Common Image Gathers}

\author{%
  Yunlin Zeng \\
  Georgia Institute of Technology \\
  \texttt{yzeng74@gatech.edu} \\
  \And
  Huseyin Tuna Erdinc \\
  Georgia Institute of Technology \\
  \And
  Felix J. Herrmann \\
  Georgia Institute of Technology \\
}

\begin{document}
\maketitle

\begin{abstract}
We present a multi-parameter simulation-based inference framework for joint
Bayesian recovery of subsurface velocity and acoustic impedance from seismic
data. A score-based diffusion model is conditioned on two complementary Common
Image Gathers (CIGs): an inverse-scattering CIG encoding reflectivity amplitude
and an anti-ISIC CIG encoding kinematic velocity errors. The model
simultaneously samples the posterior distributions of both parameters. Training
labels are deliberately decoupled to prevent the model from exploiting the
Gardner relationship: velocity targets are lightly smoothed to match the
long-wavelength content of the anti-ISIC CIG, while impedance targets retain the
unsmoothed ground truth. On the Compass benchmark the model achieves velocity
SSIM of 0.967 (RMSE 0.050 km/s) and impedance SSIM 0.867 (RMSE 0.279
km/s$\cdot$g/cm$^3$), with velocity quality confirmed by CIG focusing.
\end{abstract}


\section{Introduction}\label{introduction}

Recovering subsurface P-wave velocity and acoustic impedance
simultaneously from seismic data is central to exploration geophysics.
Velocity controls wave propagation kinematics and is the primary target
of conventional Full Waveform Inversion (FWI) \citep{virieux2009fwi},
while acoustic impedance \(Z = \rho v\) encodes rock and fluid
properties through amplitude-variation-with-offset (AVO) responses
\citep{avseth2010quantitative} and is the primary target of
multi-parameter seismic inversion. Obtaining both parameters jointly,
together with calibrated uncertainty estimates, remains a challenge.
Conventional FWI yields a single point estimate without uncertainty.
Simulation-based inference (SBI) \citep{doi:10.1073/pnas.1912789117}
addresses this by training a generative surrogate on paired physics
simulations to enable fast posterior sampling. Score-based diffusion
models are particularly effective for SBI due to their expressiveness in
capturing non-Gaussian posteriors
\citep{karras2022edm, wu2024principled}. The WISE framework
\citep{yin2024wise} demonstrated accurate velocity posteriors by
conditioning an Elucidated Diffusion Model (EDM) on subsurface-offset
Common Image Gathers (CIGs); deep learning approaches have also used CIG
volumes for velocity-model building with uncertainty quantification
\citep{geng2022deep, siahkoohi2022, orozco2024cig}. Our prior work
extended this to joint vertical and horizontal CIGs
\citep{zeng2025fwvicig}.

\begin{figure}

\centering{

\includegraphics[width=0.72\textwidth,height=\textheight]{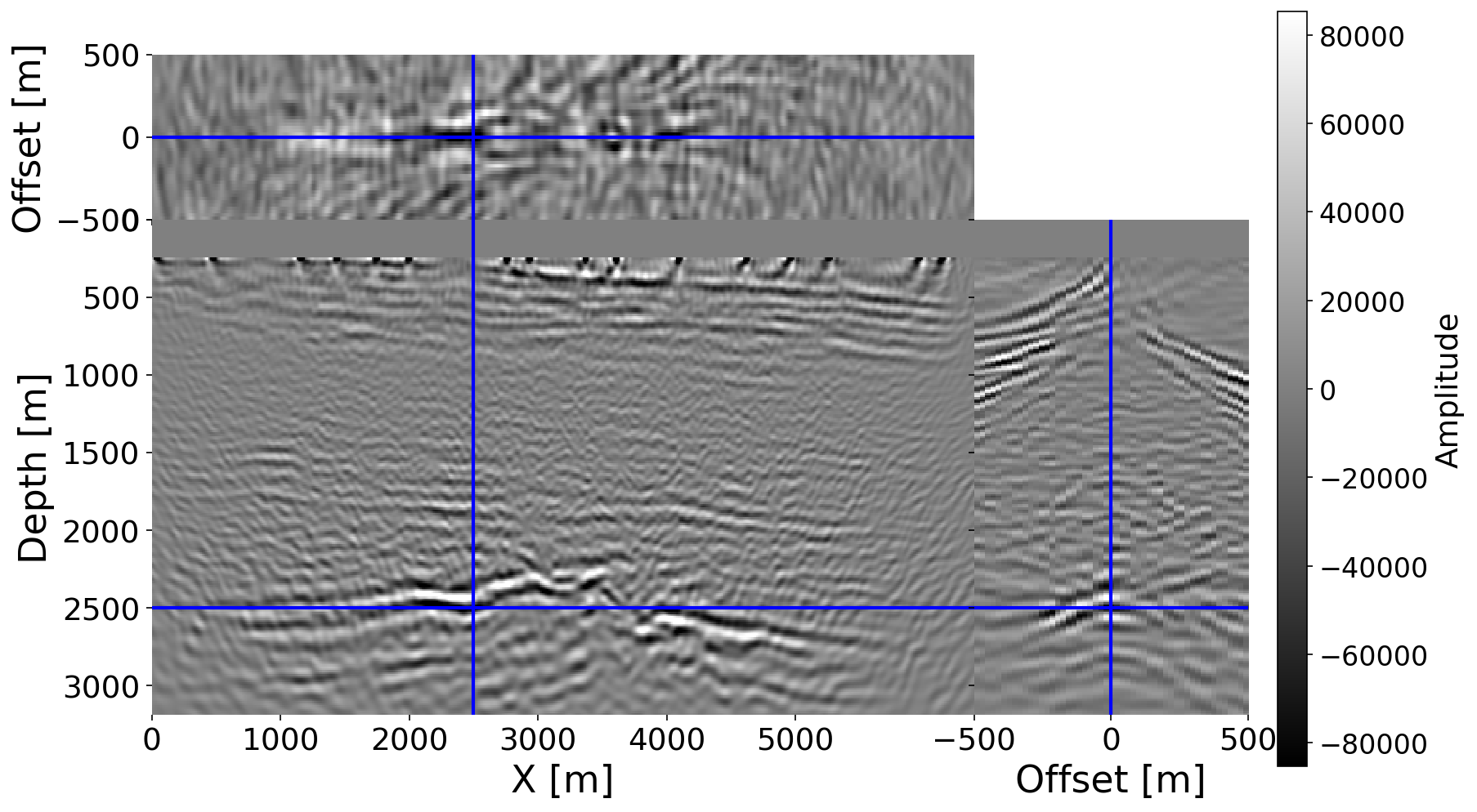}

}

\caption{\label{fig-cig-isic}ISIC CIG for a held-out test model from the
Compass benchmark. Top panel: horizontal slice at fixed depth (offset
vs.~X); bottom-left: zero-offset RTM image; bottom-right: gather at
\(x=2500\) m (depth vs.~offset). The ISIC CIG captures broadband
reflectivity amplitude, providing a short-wavelength impedance
constraint.}

\end{figure}

\begin{figure}

\centering{

\includegraphics[width=0.72\textwidth,height=\textheight]{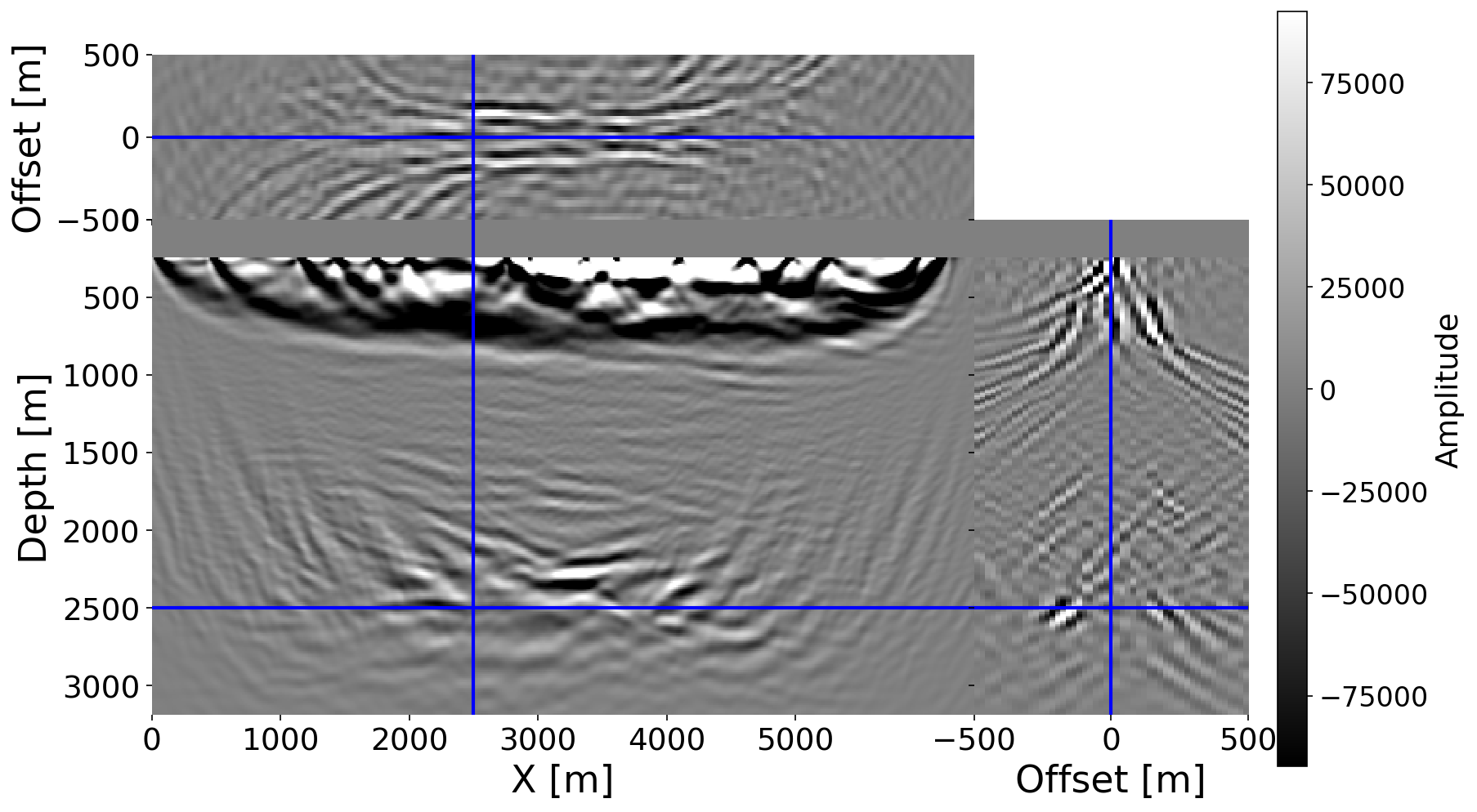}

}

\caption{\label{fig-cig-fwi}Anti-ISIC CIG for the same test model. The
gather at \(x=2500\) m (bottom-right) shows strong offset-dependent
moveout that encodes the kinematic velocity error between the smooth
background and the true velocity field.}

\end{figure}

Here we extend to simultaneous Bayesian recovery of velocity and
impedance by conditioning on two imaging conditions with fundamentally
different physics. The first imaging condition corresponds to the
Inverse Scattering Imaging Condition (ISIC,
\citep{sava2011extended, Whitmore2012, Albano2022}), which corresponds
to the gradient with respect to the acoustic impedance for constant
density. As such, this imaging condition emphasizes high-frequency
reflections at the expense of tomographic updates to the velocity model.
The second imaging condition, which we coin the anti-ISIC imaging
condition, does the opposite and brings out tomographic updates to the
velocity at the expense of reflectivity. We can observe the difference
between these two imaging conditions by juxtaposing
Figure~\ref{fig-cig-isic} and Figure~\ref{fig-cig-fwi}, which illustrate
their complementary character on a held-out test model from the Compass
benchmark.

A key challenge is the Gardner relation \citep{gardner1974formation}
(\(\rho = 0.31(1000v)^{0.25}\), \(Z=\rho v\)): a network could predict
impedance trivially via \(Z\propto v^{1.25}\) without learning ISIC
amplitude physics. We prevent this by decoupling training labels:
velocity targets are lightly smoothed (defined precisely in the
Methodology section below) to match what the anti-ISIC CIG can resolve
kinematically, while impedance targets retain the unsmoothed ground
truth via Gardner's law applied to the raw velocity, preserving sharp
reflectors. The forward simulation uses Gardner density applied to the
lightly smoothed velocity for kinematic self-consistency with the
velocity label.

\section{Methodology}\label{methodology}

Given observed shot data \(\mathbf{d}_\text{obs}\) and a smooth
background \(\mathbf{x}_0\) (Gaussian-smoothed slowness, \(\sigma{=}20\)
grid samples), two \(N_h{=}51\)-offset CIGs are computed: \[
\mathbf{y}_\text{ISIC} = \nabla \overline{\mathcal{F}}_\text{ISIC}[\mathbf{x}_0]^\top \Delta\mathbf{d}, \qquad \mathbf{y}_\text{aISIC} = \nabla \overline{\mathcal{F}}_\text{aISIC}[\mathbf{x}_0]^\top \Delta\mathbf{d},
\] where
\(\Delta\mathbf{d} = \mathcal{F}(\mathbf{x}_0) - \mathbf{d}_\text{obs}\)
is the data residual, \(\mathcal{F}(\mathbf{x}_0)\) is the acoustic
wave-equation forward operator evaluated at the background model, and
\(\nabla \overline{\mathcal{F}}_\text{ISIC/aISIC}[\mathbf{x}_0]\) is the
extended Born operator for the ISIC and anti-ISIC imaging conditions
respectively. Both CIGs are concatenated into a 102-channel conditioning
tensor \(\mathbf{y} \in \mathbb{R}^{102 \times N_z \times N_x}\)
(\(N_z{=}256\), \(N_x{=}512\)), normalized by per-channel statistics
estimated on the training set.

A lightweight summary UNet \(h_\phi\) (encoder-decoder with skip
connections) compresses \(\mathbf{y}\) from 102 channels to 18 summary
channels, discarding CIG features uninformative for inversion while
preserving the complementary velocity and impedance information. An
EDM-preconditioned SongUNet \(D_\theta\) \citep{karras2022edm} then
jointly denoises the two-channel target
\(\mathbf{x} = (\mathbf{v}, \mathbf{Z})\) conditioned on the summary: \[
\mathcal{L}(\theta,\phi) = \mathbb{E}_{\mathbf{x},\mathbf{y},\mathbf{n}} \bigl\| D_\theta\!\bigl([\mathbf{x}+\mathbf{n},\, h_\phi(\mathbf{y})],\sigma_n\bigr) - \mathbf{x} \bigr\|_2^2, \quad \mathbf{n}\sim\mathcal{N}(0,\sigma_n^2 I).
\] Here, the symbol \(\sigma_n\) denotes the noise level in the EDM
framework (distinct from the spatial Gaussian smoothing parameter used
for the velocity targets). Velocity-model training targets are obtained
by applying a Gaussian filter with standard deviation \(\sigma{=}2\)
grid samples (each sample is 12.5 m, so \(\sigma{=}25\) m) to the raw
ground-truth velocity model. At this level of smoothing, long-wavelength
structure is preserved while short-wavelength reflectivity is
suppressed, matching the kinematic resolution achievable from the
anti-ISIC CIG. We refer to these lightly smoothed models as ground-truth
velocity throughout this paper. The joint two-channel output enables the
model to learn the full multi-parameter velocity-impedance posterior,
including residual coupling not captured by the Gardner relation. At
inference, 16 posterior samples are drawn via the EDM probability-flow
ODE, providing both a posterior mean estimate and a pixel-wise
uncertainty map.

The summary UNet uses 4 encoder and 4 decoder blocks with channel
progression {[}64, 128, 256, 512{]} and a 1024-channel bottleneck. The
EDM SongUNet operates at three resolution levels (256, 128, 64 pixels)
with 4 residual blocks per level, base channel width 64, and
self-attention at the \(16{\times}16\) resolution. The combined model
totals approximately 54 million trainable parameters.

\section{Experiment Setup}\label{experiment-setup}

We use the 1000-model Compass benchmark \citep{BG} (\(512{\times}256\)
grid, \(\Delta x{=}12.5\) m) with North Sea geology including complex
structural interfaces and faults. Observed data are generated from the
lightly smoothed ground-truth velocity (\(N_\text{src}{=}16\) jittered
sources, Gardner density \citep{gardner1974formation} applied to the
lightly smoothed velocity, Ricker wavelet, SNR=12 dB), ensuring CIG
residuals are kinematically consistent with the velocity labels.
Impedance labels use the unsmoothed raw ground-truth impedance via
Gardner's law applied to the raw velocity, decoupled from the velocity
label smoothing. This forced decoupling requires the network to extract
velocity from anti-ISIC CIG kinematics and impedance from ISIC CIG
amplitude independently. Training uses model indices 1-800; validation
uses 801-1000; training runs for 100 epochs with batch size 2 and Adam
optimizer (\(\text{lr}{=}10^{-4}\)).

\section{Results}\label{results}

\begin{figure}

\centering{

\includegraphics[width=0.72\textwidth,height=\textheight]{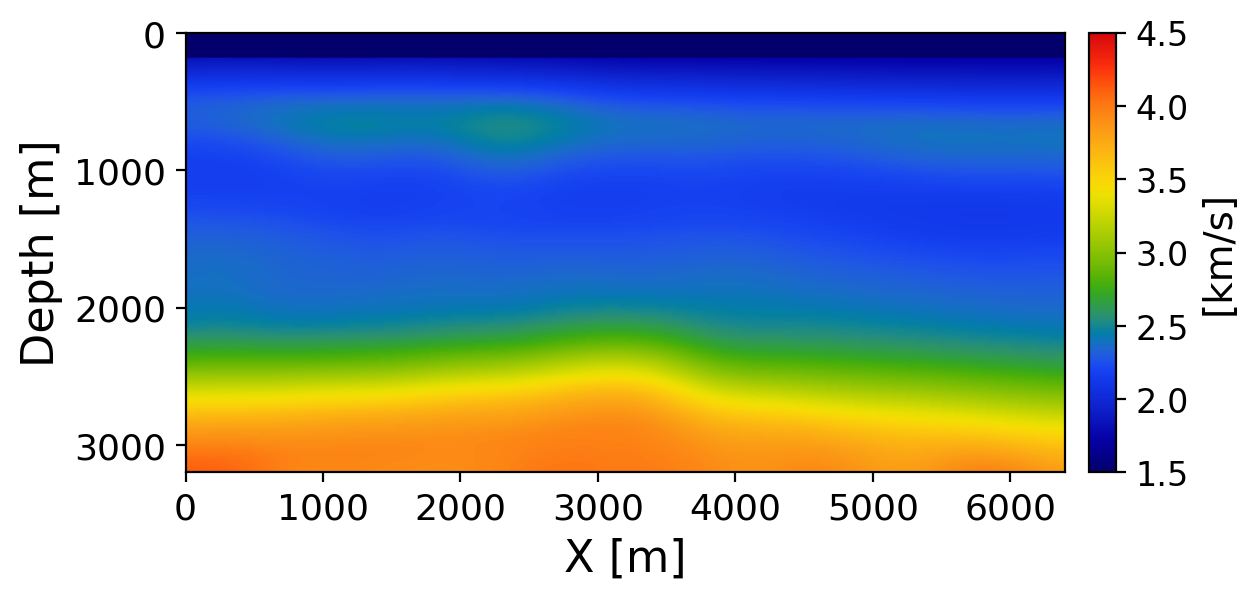}

}

\caption{\label{fig-background}Background velocity model used as
migration input for CIG computation. This model is obtained by applying
a Gaussian filter with standard deviation \(\sigma{=}20\) grid samples
to the slowness field.}

\end{figure}

\begin{figure}

\centering{

\includegraphics[width=0.72\textwidth,height=\textheight]{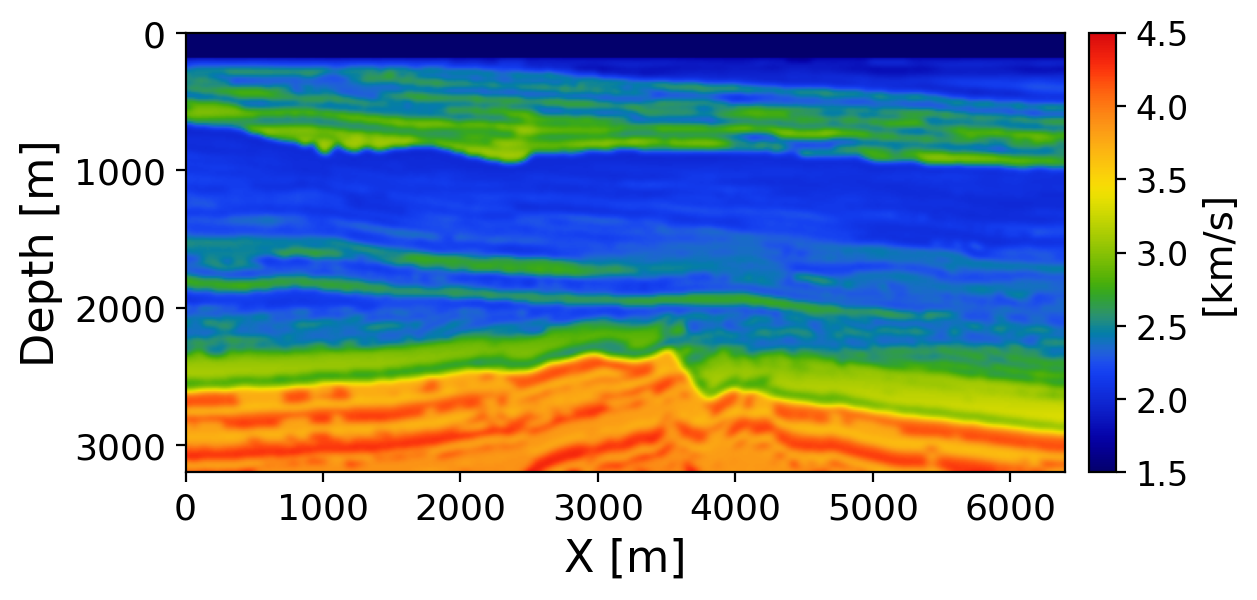}

}

\caption{\label{fig-gt-vel}Ground-truth (lightly smoothed) velocity for
a held-out test model.}

\end{figure}

We evaluate on a held-out test model from the Compass benchmark using 16
posterior samples. Figure~\ref{fig-background} shows the background
velocity used as migration input and Figure~\ref{fig-gt-vel} shows the
lightly smoothed ground-truth velocity. The background is spectrally
smooth and lacks the structural detail present in the ground truth. The
model must recover these features from the data residuals used to form
the CIGs alone.

\begin{figure}

\centering{

\includegraphics[width=0.72\textwidth,height=\textheight]{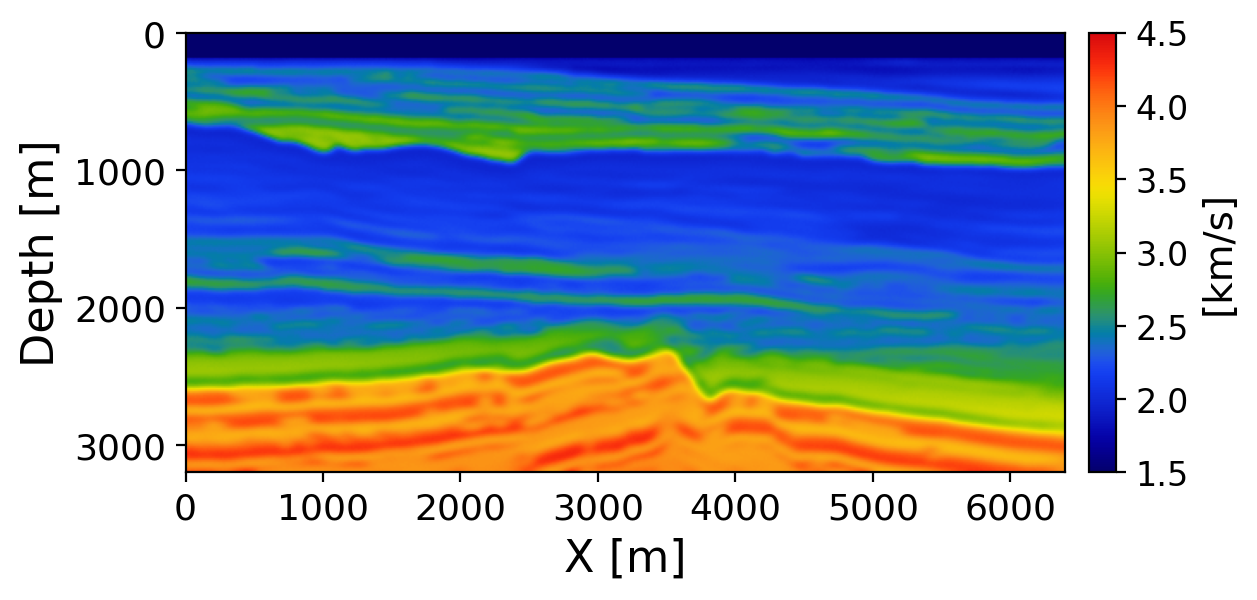}

}

\caption{\label{fig-vel}Posterior mean velocity from 16 samples.
SSIM=0.967, RMSE=0.050 km/s.}

\end{figure}

Figure~\ref{fig-vel} shows the posterior mean velocity (SSIM=0.967,
RMSE=0.050 km/s, both measured against the lightly smoothed ground-truth
velocity). Comparing with Figure~\ref{fig-gt-vel}, the model
successfully recovers the water layer, fine-scale sedimentary layering,
and deep structural interfaces from only 16 jittered sources, recovering
structure entirely absent from the smooth background.

\begin{figure}

\centering{

\includegraphics[width=0.72\textwidth,height=\textheight]{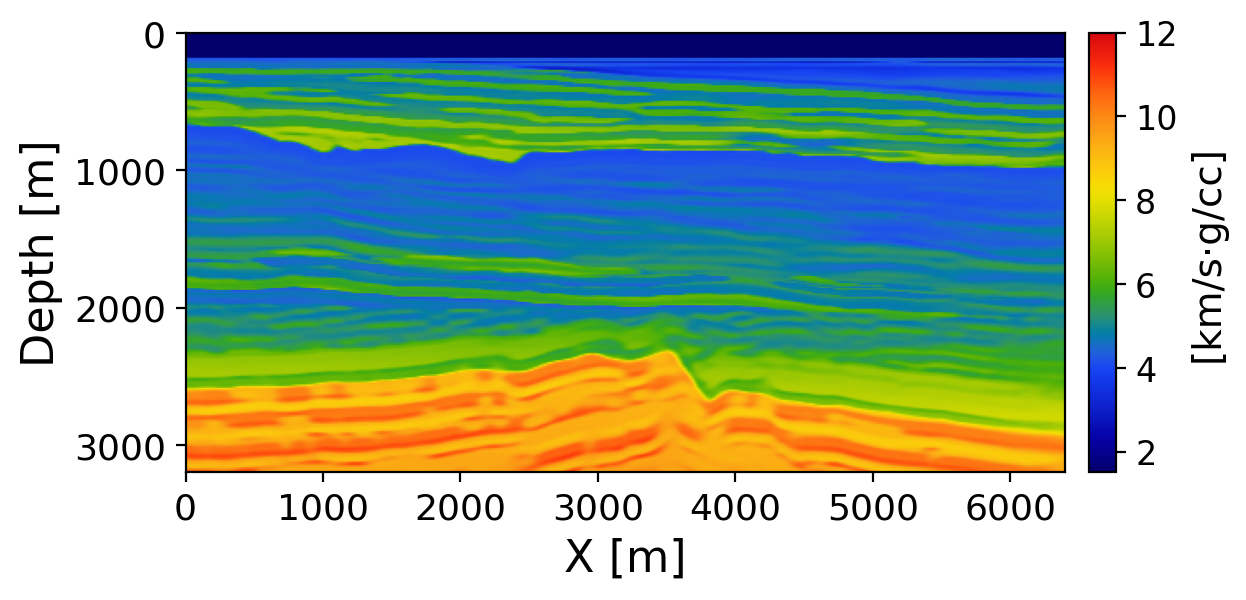}

}

\caption{\label{fig-imp}Posterior mean acoustic impedance, SSIM=0.867,
RMSE=0.279 km/s·g/cm³.}

\end{figure}

Figure~\ref{fig-imp} shows the posterior mean acoustic impedance
(SSIM=0.867, RMSE=0.279 km/s·g/cm³, measured against the ground-truth
impedance). Impedance is recovered at substantially higher spatial
resolution than velocity, resolving sharp sedimentary reflectors absent
in the smooth velocity. This confirms the model exploits the multiscale
ISIC CIG amplitude information
\citep{sava2011extended, Whitmore2012, Albano2022} rather than inferring
impedance from velocity via the Gardner relation.

\begin{figure}

\centering{

\includegraphics[width=0.72\textwidth,height=\textheight]{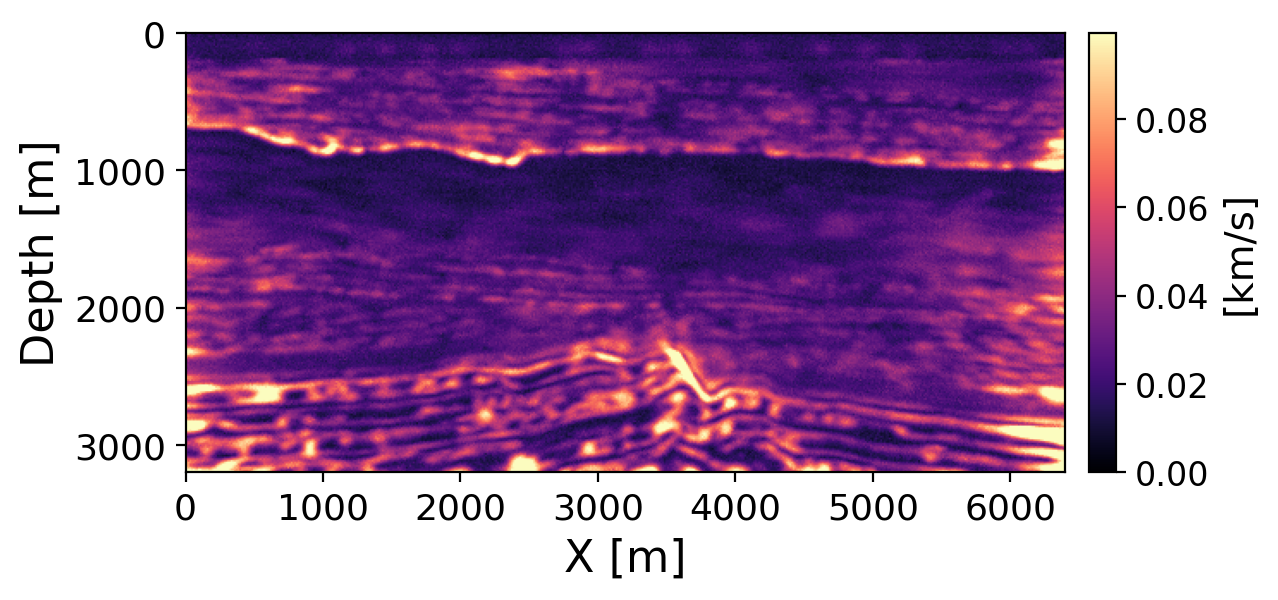}

}

\caption{\label{fig-unc-vel}Posterior standard deviation for velocity
from 16 samples.}

\end{figure}

\begin{figure}

\centering{

\includegraphics[width=0.72\textwidth,height=\textheight]{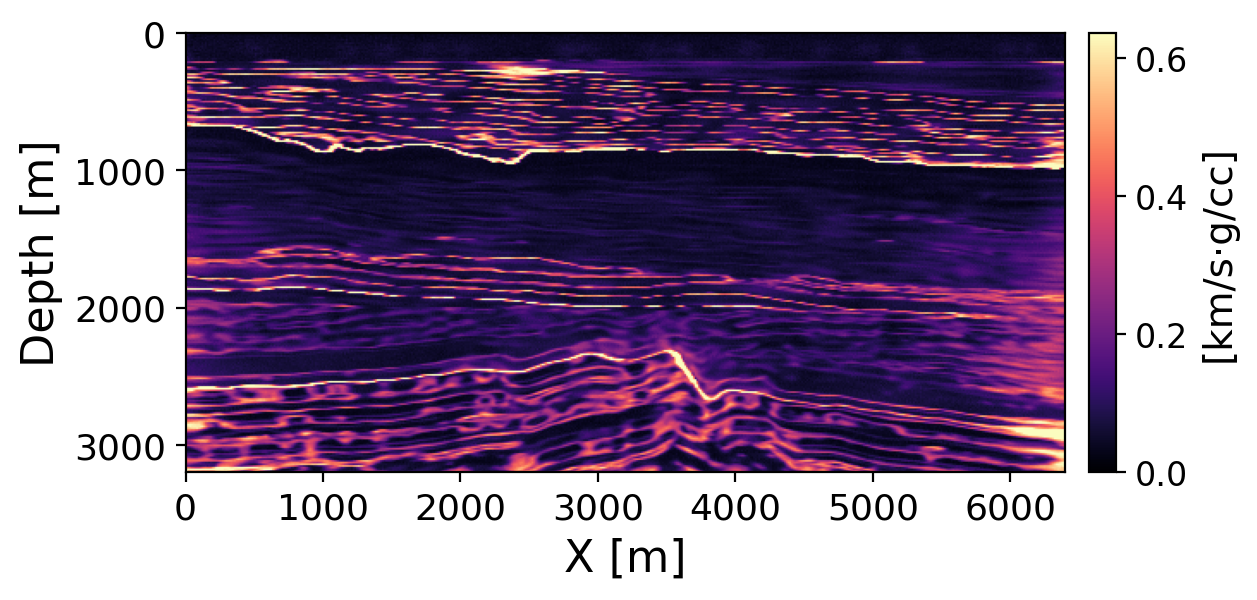}

}

\caption{\label{fig-unc-imp}Posterior standard deviation for impedance
from 16 samples.}

\end{figure}

Figure~\ref{fig-unc-vel} and Figure~\ref{fig-unc-imp} show pixel-wise
posterior standard deviation for velocity and impedance, respectively.
Velocity uncertainty is highest at structural interfaces with strong
velocity contrasts and increases with depth where sparse 16-source
acquisition limits illumination. Impedance uncertainty, by contrast,
concentrates along reflective interfaces (water bottom, intra-sediment
reflectors) rather than in broad depth zones, consistent with the ISIC
CIG sensitivity to short-wavelength amplitude contrasts
\citep{sava2011extended, kumar2013SEGAVA}.

\begin{figure}

\centering{

\includegraphics[width=0.72\textwidth,height=\textheight]{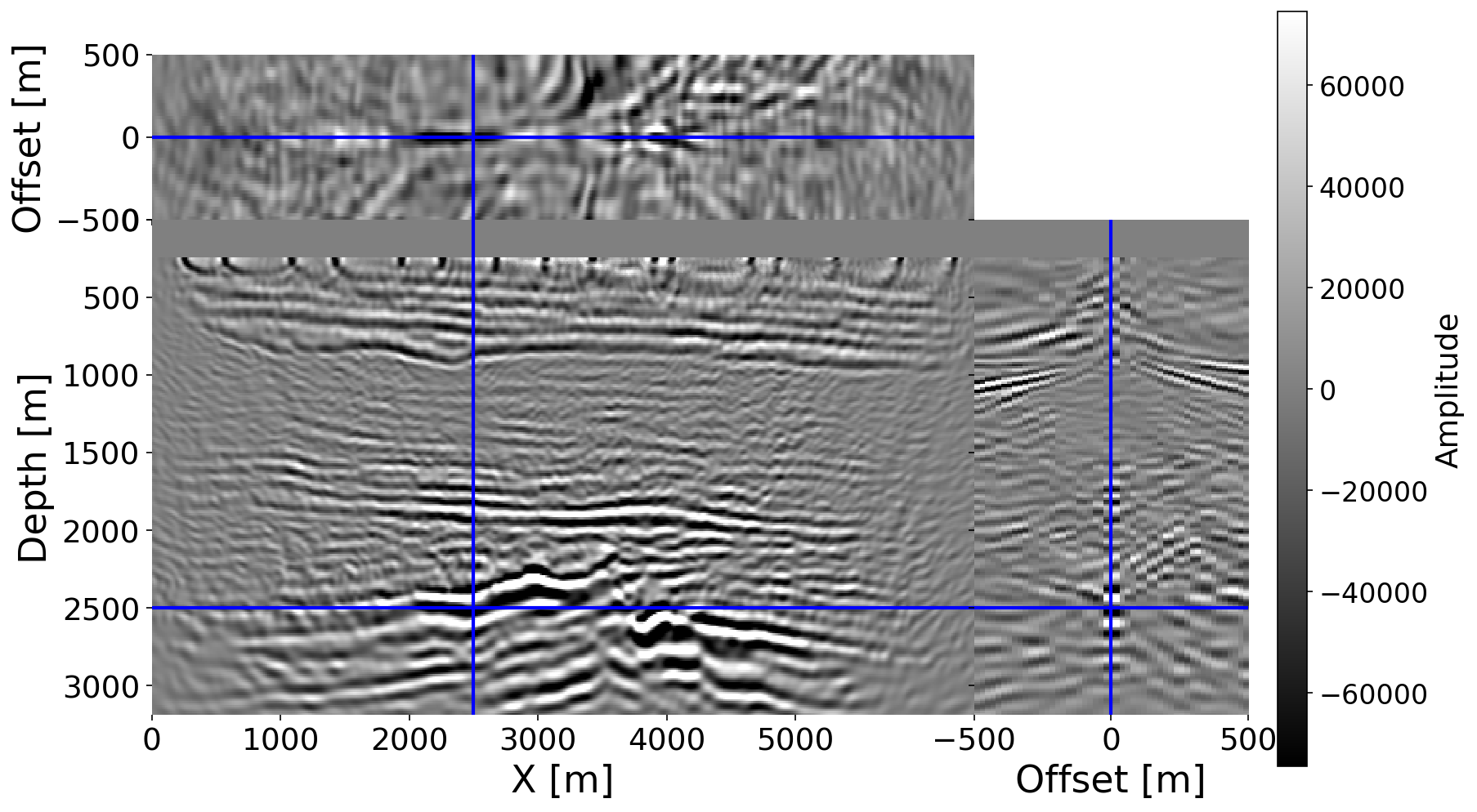}

}

\caption{\label{fig-cig-mean}ISIC CIG computed using the EDM posterior
mean velocity (conditioned on both ISIC and anti-ISIC CIGs) as migration
background.}

\end{figure}

Beyond statistical metrics, we validate the velocity quality through CIG
focusing. This comparison uses the full model conditioned on both ISIC
and anti-ISIC CIGs. Using the same lightly smoothed ground-truth
velocity as the true model, we compute new ISIC CIGs with two migration
backgrounds: the original smooth background (\(\sigma{=}20\), same as
Figure~\ref{fig-cig-isic}) and the EDM posterior mean (conditioned on
both ISIC and anti-ISIC CIGs). Figure~\ref{fig-cig-isic} already shows
the result with the smooth background: the gather at \(x{=}2500\) m
exhibits curved offset moveout indicating incorrect migration
kinematics. Figure~\ref{fig-cig-mean} shows the result with the
posterior mean background: events are substantially more focused around
zero offset and reflectors appear at their correct depths, confirming
the posterior mean velocity correctly predicts wave propagation
kinematics. This physics-based validation confirms the posterior mean is
not merely statistically accurate but constitutes a geophysically
meaningful velocity model suitable for seismic imaging.

\begin{figure}

\centering{

\includegraphics[width=0.82\textwidth,height=\textheight]{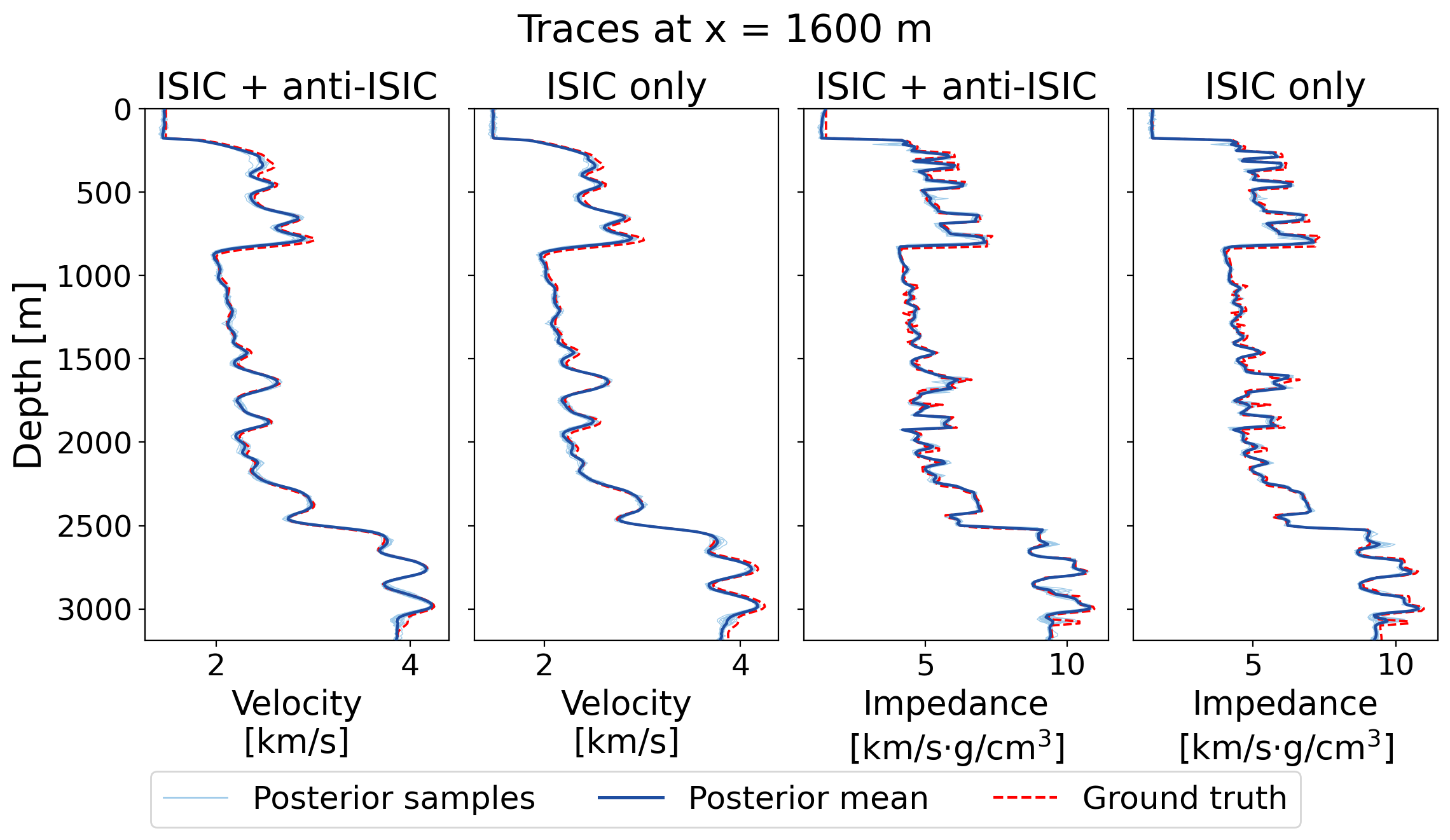}

}

\caption{\label{fig-ablation}Traces at \(x{=}1600\) m. The two left
panels show velocity and the two right panels show impedance; within each
pair, the first is the full model (ISIC + anti-ISIC) and the second is the
ISIC-only ablation. Ground truth (red dashed), posterior mean (blue solid),
individual samples (light blue).}

\end{figure}

\begin{figure}

\centering{

\includegraphics[width=0.82\textwidth,height=\textheight]{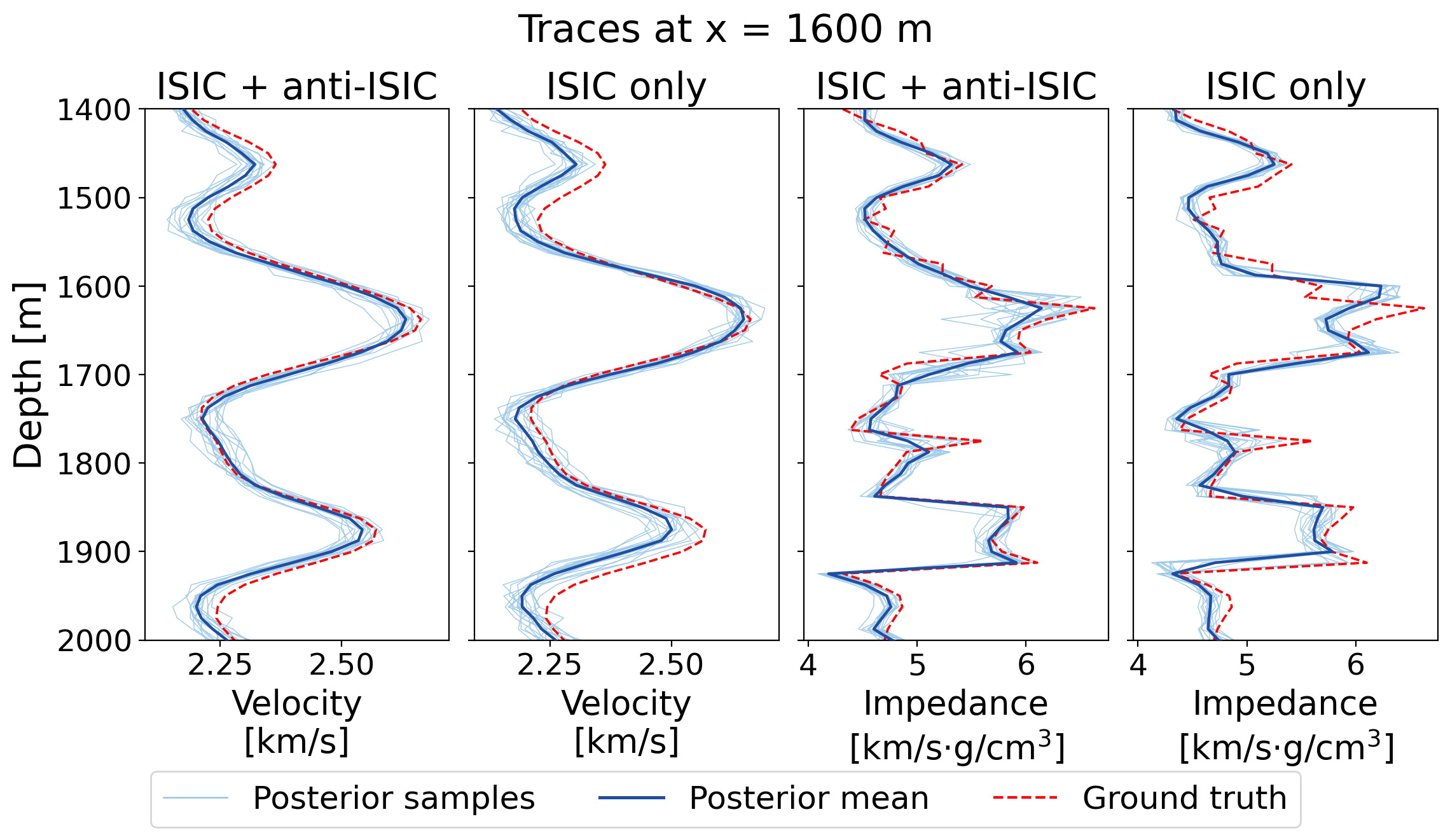}

}

\caption{\label{fig-ablation-zoom}Zoomed-in version of
Figure~\ref{fig-ablation} for the \(1400\)--\(2000\) m depth range.}

\end{figure}

We further compare the full model (ISIC + anti-ISIC, 18 summary
channels) against an ISIC-only ablation (9 summary channels).
Figure~\ref{fig-ablation} shows vertical traces at \(x{=}1600\) m for
both configurations. For velocity, conditioning on both CIG types
improves alignment with the ground truth particularly at depths
\(\geq 1000\) m, where the anti-ISIC CIG provides additional kinematic
constraints. For impedance, the full model yields better ground-truth
alignment overall at this position, suggesting the anti-ISIC CIG
contributes complementary multiscale information that benefits impedance
recovery as well. Over the full depth range the two configurations are
difficult to distinguish by eye. Figure~\ref{fig-ablation-zoom} is a zoomed-in
version for the \(1400\)--\(2000\) m range, where the model conditioned on
both CIG types aligns more closely with the ground-truth velocity near
\(1500\) m and again near \(1900\) m.

\section{Conclusions}\label{conclusions}

We presented a joint multi-parameter Bayesian inversion framework that
simultaneously recovers P-wave velocity and acoustic impedance by
conditioning an EDM diffusion model on complementary ISIC and anti-ISIC
Common Image Gathers. The two CIG types supply fundamentally different
information: the anti-ISIC CIG constrains long-wavelength tomographic
velocity model updates, while the ISIC CIG constrains short-wavelength
impedance contrasts through broadband reflectivity amplitudes.
Deliberate training label decoupling prevents the model from bypassing
amplitude physics via the Gardner relation.

On the 1000-model Compass benchmark, the posterior mean velocity
achieves SSIM=0.967 and RMSE=0.050 km/s, and the posterior mean
impedance achieves SSIM=0.867 and RMSE=0.279 km/s·g/cm³. Velocity
quality is confirmed by CIG focusing: replacing the smooth migration
background with the EDM posterior mean produces substantially more
focused events in the offset gather, demonstrating that the inferred
velocity is not only statistically accurate but geophysically meaningful
as a migration input. The joint posterior further provides calibrated,
spatially structured uncertainty: velocity uncertainty concentrates at
structural interfaces and at depth where 16-source acquisition limits
illumination, while impedance uncertainty localizes at reflective
interfaces rather than in broad depth zones.

These results suggest that conditioning diffusion models on
complementary CIG types provides a principled and computationally
efficient path toward joint parameter recovery with uncertainty
quantification in seismic exploration. Several directions remain open.
First, the current framework uses 16 sources per model; evaluating
accuracy as a function of source count will clarify the acquisition
requirements for reliable joint inversion. Second, extending the
evaluation to the full 200-sample validation set and to a range of
geological scenarios will establish the statistical robustness of the
posteriors. Third, the summary network compression ratio (102 to 18
channels) represents a design choice whose impact on posterior quality
warrants systematic ablation. Finally, adapting the framework to field
data requires bridging the simulation-to-real gap through domain
randomization or transfer learning strategies, which we identify as the
most critical next step toward practical deployment.

\section{Acknowledgement}\label{acknowledgement}

This research was carried out with the support of Georgia Research
Alliance and partners of the ML4Seismic Center. The authors acknowledge
the use of Claude and Claude Code (Anthropic) for writing assistance and
code execution.

\bibliographystyle{IMAGE2025}
\bibliography{abstract.bib}

\end{document}